\documentclass[preprint,showpacs,12pt,floatfix,aps]{revtex4}

\usepackage{graphicx}
\usepackage{latexsym}
\usepackage{amsmath}
\usepackage{subfigure}
\usepackage{upgreek}

\begin{document}

\title{Radiating Shear-Free Gravitational Collapse with Charge} 
\author{Pinheiro, G. $^{1,3}$}
\email{gpinheiro@on.br}
\author{Chan, R. $^{2}$}
\email{chan@on.br}
\affiliation{\small $^{1}$ Divis\~ao de Programas de P\'os-Gradua\c c\~ao, Observat\'orio Nacional, 
Rua General Jos\'e Cristino, 77, S\~ao Crist\'ov\~ao, 20921-400, Rio de Janeiro, RJ, Brazil
\\$^{2}$ Coordena\c{c}\~ao de Astronomia e Astrof\'{\i}sica, Observat\'orio Nacional, 
Rua General Jos\'e Cristino, 77, S\~ao Crist\'ov\~ao
20921-400, Rio de Janeiro, RJ, Brazil\\
$^{3}$ Departamento de F\'isica Te\'orica, Instituto de F\'isica, Universidade do Estado do Rio de Janeiro, Rua S\~ao Francisco Xavier, 524, Maracan\~a, CEP 20550-900, Rio de Janeiro - RJ, Brasil}

\date{\today}

\begin{abstract}
We present here a new shear free model for the gravitational collapse of a 
spherically symmetric charged body. We propose a dissipative contraction  
with radiation emitted outwards represented by the Vaidya-Reissner-Nordstr\"om metric. The Einstein field equations, using
the junction conditions and an ansatz, are integrated numerically. 
A check of the energy conditions is also performed. We obtain that the 
charge delays the Reissner-Nordstr\"om black hole formation and it can even prevent the collapse.
\end{abstract}
	
\maketitle

{\bf Keywords:}
Gravitational Collapse, General Relativity, Black Hole, Reissner-Nordstr\"om

\section{Introduction}
\label{introduction}

The equilibrium of the stars depends on the balance of two conflicting effects.
The gravitational force attracting the material of the star toward the center,
and the opposite internal thermal pressure provided by nuclear fusion of the 
elements in the stellar interior. When these reactions cease and no other 
source of pressure acts significantly, like the non-thermal degeneracy pressure, 
this balance is broken and a massive star undergoes the continuous catastrophic 
contraction, the gravitational collapse. The  remnant product of this process can 
be a black hole.

The General Theory of Relativity furnishes the foundation for 
analyzing the effects of the huge gravitational field involved in this process. 
In this context, the first idealized model for the gravitational collapse was 
proposed by Oppenheimer and Snyder \cite{OS1939}, in 1939. It consists on the 
dynamics of a simple pressureless fluid (dust), a set of no interacting particles subjected only to the action of the gravitational field generated by their own masses. In consequence, the particles travel along timelike geodesics towards the center of the configuration, resulting in the formation of a Schwarzschild black hole. Thus, an interesting step forward is to verify the consequences in considering more relevant fluids, for instance, endowed with pressure, viscosity and/or electric charge.
Since this landmark paper, many attempts aiming a more realistic description of this phenomenon have been taken place, 
but this usually impose a more complete set of equations, and the non linearity coming from the own structure of the field equations often compel us to the application of numerical procedures.

An interesting issue lies on the fact that, if a star could hold a non null 
amount of net electric charge,  the contraction of such an object could give 
rise to the Reissner-Nordstr\"om black hole.
However, the existence of considerable fraction of net charge in the stellar interior 
has been for long rejected. A process proposed by Rosseland 
\cite{Rosseland} asserts that the free electrons of the strongly ionized gas
that composes the star can be ejected due to its large thermal velocities. 
As the balance of forces is established, the process is interrupted and an object 
containing only 100 Coulomb per solar mass remains. Similar arguments 
supported by Glendenning \cite{Glendenning} reaffirm the neutrality of the 
stars. However, these objections are grounded in the Newtonian context and, as 
it was already advertised \cite{Ray03}, do not apply when extremely huge 
gravitational fields are in play. Indeed,  in the scenario of the Theory of 
the General Relativity, a relativistic star could hold greater portions of 
charge keeping itself stable (\cite{Ray03}, \cite{Ghezzi05}),
albeit no mechanism producing charge asymmetry is known. A work in this direction can be found in \cite{Cuesta03}.
Nevertheless, these issues seem far from be concluded, making it important to draw attention to the charged stellar models.

There exist several relativistic static interior charged fluid solutions connected to the exterior Reissner-Nordstr\"om spacetime. A good classification and description of most of them can be found in Ivanov, 2002 \cite{Ivanov02}. Among these solutions, the one developed by Cooperstock and De La Cruz \cite{CooperstockDeLaCruz} will be useful for our investigation. It represents a charged dust fluid, where the $T_{00}$ component of the stress-energy tensor in the comoving frame -- composed by the matter and the interior electric field -- is a constant, giving rise to a Schwarzschild-like interior solution.


In spite of the several static charged interior metrics, dynamic solutions are rare. In this sense, an interesting work performed by Bekenstein in 1971 \cite{Bekenstein71} has indicated the dual role played by the charge. Depending on the charge-mass ratio, the electric field can contribute to the support or, in opposition, to the speeding up of the structure's collapse (charge regeneration). The proof of the Birkhoff theorem and the hydrostatic equilibrium equation for the charged case were also achieved.


The Vaidya-Reissner-Nordstr\"om metric, generalization of the Vaidya's metric \cite{Vaidya53} in the presence of electromagnetic field, enables the geometrical description of the radiation field surrounding the star, allowing the development of several models of radiative gravitational collapse of charged stars \cite{deOliveira87,Medina88,MaharajGovender00,Rosales2010}. Among these models, the non adiabatic shear free model of de Oliveira and Santos \cite{deOliveira87} is very similar to the one proposed in the present paper, despite the fact that the field equations have not been integrated before and of the absence of pressure anisotropy. Some years earlier, the same authors and Kolassis \cite{deOliveira85} studied the uncharged case by means of the separation of the metric functions of the general spherically symmetric isotropic interior metric into distinct functions of the radial and temporal coordinates, a proposition already known from a preceding investigation \cite{Glass81}. Years later these models have been generalized by Chan for the shearing model \cite{Chan03}. This simplification allows the complete integration of the Einstein field equations, provided a radial function representing the static pre-collapse configuration is given. Here we use this same assumption, besides the selection of the referred Cooperstock and De La Cruz solution \cite{CooperstockDeLaCruz} for the the radial function that represents the initially static situation.
We are going to see that this ansatz forces the physical variables out of
equilibrium to have the same radial behavior as if they were in equilibrium.
This is called in the literature as the post-quasi-static approximation
\cite{HBDS2002,HB2011,B2010}.

In studying a static relativistic stellar model, Bowers \& Liang \cite{BowersLiang74} pointed to the relevant effects played by the anisotropy of pressure over the maximum mass for a stable system and over the surface redshift. 
among other factors, pressure anisotropy is induced by the electric charge \cite{Ivanov2010}, by the shear movement \cite{Chan00,Chan03} and, moreover, it is enlarged by the shear viscosity \cite{Nogueira04} (see \cite{HerreraSantos97} for a detailed review article about the occurrence of local pressure anisotropy in the study of self-gravitating systems). The shear movement, on the other hand, affects the luminosity and total collapse time of the contracting systems \cite{Chan97,Chan98a}, justifying its inclusion in several works.
However, in an unsuccessful attempt to obtain an exact solutions of the Einstein's equations, in this paper we forsake the general case of shearing collapse, although, due to others sources, the pressure anisotropy is taken into account.


Here we have to remind important consequences of the shear-free condition already studied. It was shown by Herrera and Santos \cite{HerreraSantos03,HerreraSantosWang08} that, in the quasi-static approximation and for the non-dissipative situation, the shear-free flow is equivalent to the homologous contraction in the Newtonian limit (details about homologous contraction, homology relations and its application in the astrophysics can be found in \cite{Kippenhahn}). Moreover, considering dissipation in both regimes, diffusion and streaming out, the authors verified the need to impose homology conditions on the temperature and emission rate to keep the homologous evolution \cite{HerreraSantos03}. In addition, it was checked that the requirement of homogeneous expansion implies homology conditions.
Another important aspect is that the shear-free condition is not assured all along the collapse of spheres whether the dissipative processes, the density inhomogeneity and the pressure anisotropy are taking in account \cite{HerreraDiPriscoOspino}. Hence, this follows, leastwise in the study of geodesic motion of the fluid.


Current literature is comprised by some comprehensive models of charged collapsing radiating spheres with shear motion.
A good example is the work developed by Medina and collaborators \cite{Medina88} with the aid of a method developed by Herrera et. al. \cite{Herrera80}. 
Following the procedure, a system of first order differential equations evaluated at the surface is obtained and integrated. The same method was applied by Rosales et. al. \cite{Rosales2010} in more recent paper. 
Treating the same subject, but using the approach of Misner and Sharp \cite{misnersharp64}, Di Prisco and collaborators \cite{DiPrisco07} analyzed the dynamics of the charged spheres.


The stellar contraction must be a highly dissipative process due mainly to the interaction of the radiation and matter in the interior, such that, in the same way we are considering emission of energy across the stellar surface, an outward flux of energy through the body occurs. The energy transport is often considered in both regimes, diffusion and free-streaming. Once the studies of the radiation of the type-II supernova 1987A pointed to the domain of the former \cite{Lattimer88}, this is the only one taken into account in this work. Details on how the energy flux happens and how it relates to the temperature are not the focus here. This could be performed, at first glance, with the usage of the relativistic thermodynamic theory developed by Eckart \cite{Eckart}. But, although simple, this theory has an undesirable feature, the equations for the heat transfer and viscous tensions indicate instantaneous propagation of the perturbations, generating causality problems. This difficulty was overcome with the Israel-Stweart theory \cite{IsraelStewart79}, where the relaxation times for the dissipative quantities were included, resulting in hyperbolic equations for the perturbations (a complete discussion can be found in \cite{Maartens96}). The energy transport equations of this theory supplied the thermodynamic analysis in several works. In an interesting paper, Di Prisco and collaborators \cite{DiPriscoHerreraEsculpi96} compared the luminosity profile of the collapsing spheres for the vanishing (Eckart) and non-vanishing (Israel-Stweart) relaxation times. The same idea was used in a different model by the same authors and some more collaborators \cite{DiPriscoetal1997} one year later. In this way, some effects of the pre-relaxation process were confirmed to be model-independent (more on the influence of the relaxation time in the outcome of the collapsing systems in \cite{HerreraMartinez98}). In applying the transport equation to collapsing sphere models, the authors usually simplify the problem making null the constants coupling heat and viscosity. Few years ago, Herrera and collaborators \cite{Herrera09}, using the the method by Misner \& Sharp \cite{misnersharp64}, studied the dissipative gravitational collapse when this simplification is set aside. In this way, the authors have appreciated the influence of these coupling constants over the effective inertial mass of the system.


The aim of the present investigation is to find a dynamical solution for a charged sphere representing the 
dissipative collapse of the charged body, as well as to verify the role played by the 
electric charge in this process. The formation or avoidance of the black hole is discussed.
The paper is organized as follows, in Section \ref{equations} we set the metrics and the stress-energy tensor used, we show the field equations and obtain a set of equations by the application of the junction conditions. The solution of the system of 
equations is performed in Section \ref{solution_equations}. The equations for 
the static initial configuration are presented in Section \ref{est_config}. 
The results can be found in Section \ref{results} and the agreement with the energy conditions are checked the Section \ref{energyconditions}. Finally, we discuss the results in Section \ref{discussion}.

\section{Equations}
\label{equations}

The system is composed by an interior spacetime (referred by a minus signal "-"), 
a comoving timelike hypersurface (referred by $\Sigma$) and an 
exterior spacetime representing a radiation field surrounding the body (referred 
by a plus sign "+").  The unknown interior spacetime is given by the spatially isotropic 
spherically symmetric shear-free metric in comoving coordinates,
\begin{equation}
ds^2_{-} = -A^2(r,t)dt^2+B^2(r,t) \left[ dr^2 + r^2(d\theta^2+\sin^2 \theta d\phi^2) \right].
\label{eq:dsi}
\end{equation}

The exterior spacetime is described by Vaidya-Reissner-Nordstr\"om metric, which represents an outgoing 
radial flux of radiation around a charged spherically symmetric source of gravitational 
field, given by
\begin{equation}
\label{vaidya-reissner-nordstrom}
ds^{2}_{+} = - \left[ 1 -\frac{2 m(v)}{\textbf{r}}+\frac{Q^2}{\textbf{r}^2} \right] dv^2 - 
2 dv d\bf{r} + \bf{r}^2 (d\theta^2+\sin^2 \theta d\phi^2), 
\end{equation}
where $m(v)$ represents the mass, function of the retarded time $v$, and $Q$ is the 
total amount of charge of the system inside the boundary surface $\Sigma$.

The interior stress-energy tensor of a charged dissipative anisotropic fluid is given by
\begin{eqnarray}
T_{\alpha \beta}&=&  (\mu+p_\bot)u_{\alpha}u_{\beta}+p_\bot g_{\alpha \beta}+ 
(p-p_\bot)X_{\alpha}X_{\beta} + \nonumber \\
& &  +q_{\alpha}u_{\beta}+q_{\beta}u_{\alpha} + \frac{1}{4 \pi} \left[F_{\alpha}^{\gamma} F_{\beta \gamma} - 
\frac{1}{4} g_{\alpha \beta} F^{\gamma \delta} F_{\gamma \delta} \right].
\label{tensorgeral}
\end{eqnarray}
These quantities are the energy density $\mu$, the radial pressure $p$, the tangential pressure $p_\bot$, the radial heat flux $q^\alpha$, $u^\alpha$ is the four-velocity and $X^{\alpha}$ is a radial 
four-vector. The following relations hold: $u_\alpha q^\alpha = 0, X_\alpha X^\alpha = 1, 
X_\alpha u^\alpha = 0$. The tensor $F_{\alpha \beta}$ represents the electromagnetic field tensor. At last, the coupling constant in geometrized units is $\kappa = 8 \pi$ (i.e., $c = G = 1$). 

As we use comoving coordinates, we have
\begin{equation}
u^{\alpha}=A^{-1}\delta^{\alpha}_0,
\label{eq:u}
\end{equation}
and as the heat flux is radial
\begin{equation}
q^{\alpha}=q\delta^{\alpha}_1.
\label{eq:qa}
\end{equation}

The Maxwell's equation are written as
\begin{equation}
F_{\alpha \beta} = \phi_{\beta,\alpha} - \phi_{\alpha,\beta},
\label{maxeq1}
\end{equation}
and
\begin{equation}
F^{\alpha \beta}_{\hspace{10pt};\hspace{1pt}\beta} = 4 \pi J^\alpha,
\label{maxeq2}
\end{equation} 
where $\phi_{\alpha}$ is the four-potential and $J^\alpha$ is the four-current. 
It is assumed that the charge is at rest with respect to the coordinates of the metric 
(\ref{eq:dsi}), so we have no magnetic field present and, thus, we can write
\begin{equation}
\phi_{\alpha}=\Phi \delta^{0}_{\alpha},
\label{4potential}
\end{equation}
and
\begin{equation}
J^{\alpha}=\sigma u^\alpha.
\label{4current}
\end{equation}
\noindent Where $\Phi$ is the scalar potential and $\sigma$ is the charge density.

From equations (\ref{4potential}) and (\ref{maxeq1}) we obtain
\begin{equation}
F_{01}=-F_{10}=-{{\partial \Phi} \over {\partial r}}.
\label{F01}
\end{equation}

Substituting equations (\ref{4current}) and (\ref{F01}) into Maxwell's equations 
(\ref{maxeq1}) and (\ref{maxeq2}), we have
\begin{equation}
\Phi ''+ \left( -\frac{A'}{A} + \frac{B'}{B} +\frac{2}{r} \right) \Phi ' = 4\pi AB^2 \sigma ,
\end{equation}
and
\begin{equation}
\frac{\partial}{\partial t} \left( \frac{\Phi '}{A^2 B^2} \right)+ 
\left( -\frac{\dot{A}}{A} + 3\frac{\dot{B}}{B} \right) \frac{\Phi '}{A^2 B^2} = 0,
\end{equation}
where the dot represents differentiation with respect to $t$ and the prime 
with respect to $r$.

Integrating these equations, we obtain
\begin{equation}
\Phi '=\frac{A}{B} \frac{l}{r^2},
\end{equation}
where $l(r)$ is the radial distribution of charge given by
\begin{equation}
l(r)= 4\pi \int \limits_{0}^{r} \sigma B^3 r^2 dr.
\end{equation}
The integration from the center ($r=0$) till the surface ($r=r_{\Sigma}$) 
provides us the total amount of charge $Q$ inside the body.

The non-vanishing components of the field equations applied to the interior 
spacetime are very similar to the  paper \cite{deOliveira87}, except by the presence -- justified in the first section -- of the anisotropy of pressure. These are obtained with the aid of the equations (\ref{eq:dsi}), (\ref{tensorgeral}), (\ref{eq:u}), (\ref{eq:qa}), resulting in
\begin{eqnarray}
G^{-}_{00} &=& - \left( {A} \over {B} \right)^2 \left[ 2 {B'' \over B} -
 \left( {B'} \over {B} \right)^2 + {4 \over r} {B'\over B} \right] + 
3 \left( {\dot{B}} \over {B} \right)^2 = \kappa \left( A^2 \mu + 
{{1} \over {8 \pi}} {A^2 \over B^4} {l^2 \over r^4} \right),
\label{eq:g00}
\end{eqnarray}
\begin{eqnarray}
G^{-}_{11} &=&  \left( {{B'} \over {B}} \right)^2 + {{2} \over {r}} {{B'} \over {B}} + 
2 {{A'} \over {A}} {{B'} \over {B}} + {{2} \over {r}} {{A'} \over {A}} + 
{ \left( B \over A \right) }^2 \left[ -2 {\ddot B \over B} - { \left( \dot B \over B \right) }^2 + 
2 {\dot A \over A} {\dot B \over B} \right] \nonumber \\
& &= \kappa B^2 \left( p - {{1} \over {8 \pi}} {1 \over B^4} {l^2 \over r^4} \right),
\label{eq:g11}
\end{eqnarray}
\begin{eqnarray}
G^{-}_{22} &=& \frac{1}{sen^2\theta}G^{-}_{33} = \left[ \frac{B''}{B} - 
\left(\frac{B'}{B}\right)^2 + \frac{1}{r} \frac{B'}{B} + \frac{A''}{A} +
\frac{1}{r} \frac{A'}{A} \right] \nonumber \\
& & + \frac{B^2}{A^2} \left[ -2\frac{\ddot{B}}{B} - \left(\frac{\dot{B}}{B} \right)^2 + 
2\frac{\dot{A}}{A} \frac{\dot{B}}{B} \right] = \kappa B^2 \left( p_\bot  + 
{{1} \over {8 \pi}} {1 \over B^4} {l^2 \over r^4} \right),
\label{eq:g22}
\end{eqnarray}
\begin{eqnarray}
G^{-}_{01} &=& -2{\dot{B'} \over B} + 2\frac{B'}{B} \frac{\dot{B}}{B} + 
2{A' \over A}{\dot B \over B}= -\kappa A B^2 q.
\label{eq:g01}
\end{eqnarray}

We can note that the equations presented in this section are a particular case of those present in the section II of the paper \cite{DiPrisco07}. Here we are assuming no shearing motion and we neglect the shear viscosity as well as the dissipation in the free-streaming regime.


We consider a spherical surface with its motion described by a timelike 
three-surface $\Sigma$, which splits spacetimes into interior and exterior manifolds. 
The connection between these geometries has to be continuous and smooth, assured 
by the junction conditions. Here we follow the approach given by \cite{Israel66a}, 
\cite{Israel66b} and demand the continuity of the metric $
(ds^2_{-})_{\Sigma}=(ds^2_{+})_{\Sigma}=ds^2_{\Sigma}$ and of the the extrinsic curvature $K^{-}_{ij}=K^{+}_{ij}$ across the surface $\Sigma$ (details on the junction conditions used here can be seen in the paper \cite{deOliveira87}).

The condition of continuity of the metric leads us to
\begin{equation}
{dt \over {d \tau}}=A(r_{\Sigma},t)^{-1},
\label{eq:ts}
\end{equation}
\begin{equation}
r_\Sigma B(r_{\Sigma},t)={{\bf r}_{\Sigma}(v)},
\label{eq:cs}
\end{equation}
and
\begin{equation}
\left( dv \over {d \tau} \right)^{-2}_{\Sigma}=\left( 1 - {2m \over {\bf r}} + 
2 {d{\bf r} \over dv} + \frac{Q^2}{\textbf{r}^2} \right)_{\Sigma},
\label{eq:dvdtau}
\end{equation}
where $\tau$ is a proper time defined on the comoving hypersurface $\Sigma$.

The non-vanishing extrinsic curvature components are $K^{\pm}_{\tau \tau}$, $K^{\pm}_{\theta \theta}$ e $K^{\pm}_{\phi \phi}$.

Equating $K^{-}_{\theta \theta}$ and $K^{+}_{\theta \theta}$ we have
\begin{equation}
\left[ {\left( dv \over d \tau \right)}
\left( 1 - {2m \over {\bf r}} -\frac{Q^2}{\textbf{r}^2} \right){\bf r}+
{d{\bf r} \over d \tau }{\bf r}
\right]_{\Sigma}= \left[ r\left(r B' + B \right) \right]_{\Sigma}.
\label{eq:k22ik22o}
\end{equation}

With the help of equations (\ref{eq:ts}), (\ref{eq:cs}), (\ref{eq:dvdtau}), 
we can write equation (\ref{eq:k22ik22o}) as
\begin{equation}
m=\left[ \frac{r^3 B\dot{B}^2}{2A^2} -r^2B'-\frac{r^3B'^2}{2B} +
\frac{Q^2}{2rB} \right]_{\Sigma},
\label{eq:ms}
\end{equation}
which is the total energy entrapped inside the surface $\Sigma$ \cite{Cahill70}.

Equating $K^{-}_{\tau \tau}$ and $K^{+}_{\tau \tau}$, using equation (\ref{eq:ts}),
we have
\begin{equation}
\left[ {d^2v \over {d \tau^2}} 
{\left( dv \over d \tau \right)}^{-1}
-{\left( dv \over d \tau \right)} {m \over {{\bf r}^2}} +\frac{Q^2}{\textbf{r}^3} 
\frac{dv}{d\tau} \right]_{\Sigma}
=-\left( {A' \over {AB}} \right)_{\Sigma}.
\label{eq:k00ik00o}
\end{equation}

Substituting equations (\ref{eq:ts}), (\ref{eq:cs}) and ({\ref{eq:ms}) 
into (\ref{eq:k22ik22o}) we can write
\begin{equation}
\left( {dv \over {d \tau}} \right)_{\Sigma}= 
\left[ { \left(rB\right)' \over B} + { {\left( rB \right)}^{\cdot} \over A} \right]^{-1}_{\Sigma}.
\label{eq:dvdtau1}
\end{equation}
This is the gravitational redshift for the observer at rest at the infinity. Its divergence indicates the formation of an event horizon. This happens when the factor in parentheses goes to zero.

Substituting the derivative of this last equation and equation (\ref{eq:ms}) into 
(\ref{eq:k00ik00o}), identifying the geometrical terms with the field equations 
(\ref{eq:g11}) and (\ref{eq:g01}), we have
\begin{equation}
p_{\Sigma}=(qB)_{\Sigma}.
\label{eq:pqbs}
\end{equation}

This result is the same of the one obtained by de Oliveira et. al. \cite{deOliveira85} and years later in the study of charged spheres \cite{deOliveira87} and \cite{MaharajGovender00}.

We can use the equations (\ref{eq:ms}), (\ref{eq:ts}),(\ref{eq:dvdtau1}) and (\ref{eq:g11}) to obtain the total luminosity for an observer at rest at infinity
\begin{equation}
L_{\infty}=-\left({dm \over dv}\right)_{\Sigma}={\kappa \over 2}\left[ p r^2 B^2 {\left( 1 + {rB' \over B} + 
{r \dot B \over \ A} \right)}^2 \right]_{\Sigma}.
\label{eq:lsp}
\end{equation}

\section{Solution of the Field Equations}
\label{solution_equations}

In order to integrate the field equations, we resort to the same ansatz referred in the Section \ref{introduction}, consisting in separate the metric functions $A(r,t)$ and $B(r,t)$ into functions of the 
coordinates $r$ and $t$ in the form
\begin{equation}
A(r,t)=A_0(r),
\label{eq:art0}
\end{equation}
and
\begin{equation}
B(r,t)=B_0(r)f(t),
\label{eq:brt0}
\end{equation}
where $A_0(r)$ and $B_0(r)$ are solutions of a static charged dust fluid.

We have chosen this separation of variables in the metric functions in such way that 
when $f(t \rightarrow -\infty) \rightarrow 1$, the metric functions represent the initial static solution. 
The decreasing (or increasing) of $f(t)$ provides the decreasing (or increasing) of the 
physical radius ($R=rB(r,t)$) of the body, describing the collapse (or expansion). 
The process ends when $f(t\rightarrow 0) \rightarrow 0$.

Now the equations (\ref{eq:g00})-(\ref{eq:g01}) can be written as
\begin{eqnarray}
\kappa \mu = \kappa \frac{\mu_0}{f^2} + \frac{l^2}{r^4{B_0}^4 f^2} \left( 1 - 
\frac{1}{f^2} \right) + \frac{3}{{A_0}^2} \left( \frac{\dot{f}}{f} \right)^2,
\label{eq:mu}
\end{eqnarray}
\begin{eqnarray}
\kappa p = \kappa \frac{p_0}{f^2} - \frac{l^2}{r^4{B_0}^4 f^2} \left( 1 - \frac{1}{f^2} \right) + 
\frac{1}{{A_0}^2} \left[ -2 \frac{\ddot{f}}{f} - \left( \frac{ \dot{f}}{f} \right)^2 \right],
\label{eq:p}
\end{eqnarray}
\begin{eqnarray}
\kappa p_{\bot} = \kappa \frac{p_{0 \bot}}{f^2} + \frac{l^2}{r^4{B_0}^4 f^2} \left( 1 - \frac{1}{f^2} \right) + \frac{1}{{A_0}^2} \left[ -2 \frac{\ddot{f}}{f} - \left( \frac{ \dot{f}}{f} \right)^2 \right],
\label{eq:pt}
\end{eqnarray}
\begin{eqnarray}
\kappa q = - \frac{2 A_0'}{{A_0}^2 {B_0}^2} \frac{ \dot{f}}{f^3},
\label{eq:q}
\end{eqnarray}
where
\begin{equation}
\kappa \mu_0 = -{1 \over {B^2_0}}\left[ 2{B''_0 \over B_0} - 
{\left( B'_0 \over B_0 \right)}^2 + {4 \over r}{B'_0 \over B_0} \right]-
\frac{l^2}{r^4 B_{0}^{4}},
\label{eq:mu0}
\end{equation}

\begin{equation}
\kappa p_0 = {1 \over {B^2_0}}\left[ {\left( B'_0 \over B_0 \right)}^2 + 
{2 \over r}{B'_0 \over B_0} + 2{A'_0 \over A_0}{B'_0 \over B_0} +
{2 \over r}{A'_0 \over A_0} \right]+\frac{l^2}{r^4 B_{0}^{4}},
\label{eq:p0}
\end{equation}
and
\begin{equation}
\kappa p_{0 \bot}= {1 \over {B^2_0}} \left[ \frac{B''_0}{B_0} -
\left( \frac{B'_0}{B_0}\right)^2 + \frac{1}{r}{B'_0 \over B_0} + {A''_0 \over A_0} + 
\frac{1}{r} \frac{A'_0}{A_0} - \frac{l^2}{r^4 B_{0}^{2}}  \right],
\label{eq:pt0}
\end{equation}

Substituting equations (\ref{eq:p}), (\ref{eq:q}) into (\ref{eq:pqbs}) and 
assuming also that $p_0(r_{\Sigma})=0$, we obtain a second order differential 
equation for $f(t)$,
\begin{equation}
2 f^3 \ddot{f} + f^2 \dot{f}^2 - 2 \overline{a} f^2 \dot{f} - \overline{b} 
\left( 1-f^2 \right) = 0,
\label{eqdiff}
\end{equation}
\noindent where
\begin{equation}
\overline{a} = \left(A_0' \over B_0 \right)_{\Sigma},
\label{eq:as}
\end{equation}
\noindent and
\begin{equation}
\overline{b} = \left( {A^2_0 Q^2 \over {r^4B^4_0}} \right)_{\Sigma}.
\label{eq:bs}
\end{equation}
This equation is similar to the one obtained by de Oliveira et al. in 1985 \cite{deOliveira85}, except by 
the last term.  The attempts to solve it analytically proved to be unfortunate, at last, the numerical solution was successful. 
In order to verify the electric field action, different values of total charge were considered.

\section{Model of the Initial Configuration}
\label{est_config}

We consider that the system at the beginning of the collapse has a static configuration of a charged dust fluid described in Cooperstock e De La Cruz \cite{CooperstockDeLaCruz}. 
The Schwarzschild-like interior solution is obtained considering

\begin{eqnarray}
T_0^0=\lambda+\frac{E^2}{8 \pi}=\frac{3a^2}{8 \pi},
\end{eqnarray}

\noindent where $T_0^0$ is the mass-energy density component of the stress-energy tensor, $\lambda$ is the mass density, $E$ is the electric field and $a$ is a constant.

The metric for this case, presented here in isotropic coordinates, is given by
\begin{eqnarray}
ds^2_{-} = A^2_0({\bar r}) dt^2 +  B^2_0({\bar r}) 
\left[ d{\bar r}^2 + {\bar r}^2 (d\theta^2+\sin^2 \theta d\phi^2) \right],
\label{coopiso}
\end{eqnarray}
where
\begin{equation}
\label{A0r}
A^2_0({\bar r})=\frac{\beta^2}{4} (1+ {\bar r}^2)^2,
\end{equation}
\begin{equation}
\label{B0r}
B^2_0({\bar r})=\frac{4}{a^2 C^2 (1+{\bar r}^2)^2},
\end{equation}
\begin{equation}
\bar r=r/C,
\end{equation}
\noindent and the constants $\beta$ and $a$ are given by
\begin{align*}
\beta = 1 - a^2 R^2 + (1 - a^2 R^2)^{1/2},
\end{align*}
\begin{align*}
a^2 = \frac{2 m_0}{R^3} - \frac{Q^2}{R^4}. \nonumber
\end{align*}
The quantities $m_0$, $R=(r_\Sigma B_0(r_\Sigma))$ and $Q$ are the total mass, radius 
and charge of the object. The constant $C$ comes from the 
integration of the original Schwarzschild coordinates transformed into the isotropic coordinates. It can be easily determined by junction conditions.

The coefficients (\ref{eq:as}) and (\ref{eq:bs}) become
\begin{eqnarray}
\label{coeficientes}
\overline{a}=\frac{a^2 \beta R}{\left( 1 + \sqrt{1-a^2 R^2}\right)^2} \\ \nonumber
\overline{b}=\frac{{\beta}^2 Q^2}{ R^4 \left( 1+\sqrt{1-a^2 R^2} \right)^2}.
\end{eqnarray}

Substituting the functions $A_0(\bar r)$, $B_0(\bar r)$, their derivatives and a rewritten version of the 
equations (\ref{eqdiff}) into the equations (\ref{eq:mu}) - (\ref{eq:p0}), we obtain

\begin{equation}
\kappa \mu = \frac{\kappa \mu_0}{f^2}+\frac{l^2 a^4 }{16{\bar r}^4 f^2} 
\left(1+{\bar r}^2 \right)^4 \left(1-\frac{1}{f^2}\right)+\frac{12}{\beta^{2}f^2
\left(1+{\bar r}^2 \right)^2}\dot{f}^2,
\label{eqdensidade}
\end{equation}

\begin{equation}
\kappa p = -\frac{l^2 a^4 }{16{\bar r}^4 f^2} \left (1+{\bar r}^2 \right)^4 
\left(1-\frac{1}{f^2}\right)+\frac{4}{\beta^{2} f^2 \left (1+{\bar r}^2 \right)^2} 
\left[\overline{b}\left(1-\frac{1}{f^2}\right) -2\overline{a} \dot{f} \right],
\label{eqpressao}
\end{equation}

\begin{equation}
\kappa p_\bot = \frac{l^2 a^4 }{16{\bar r}^4 f^2} \left (1+{\bar r}^2 \right)^4 
\left(1-\frac{1}{f^2}\right)+\frac{4}{\beta^{2} f^2 \left (1+{\bar r}^2 \right)^2} 
\left[\overline{b}\left(1-\frac{1}{f^2}\right) -2\overline{a} \dot{f} \right],
\label{eqpressaot}
\end{equation}

\begin{equation}
\kappa q=-\frac{2a^2C{\bar r}}{\beta} \frac{\dot{f}}{f^3},
\label{eqcalor}
\end{equation}
where
\begin{eqnarray}
& &\mu_0({\bar r}) = \frac{a^2 \left(3-{\bar r}^2\right)}{8\pi}, \\
& & p_0({\bar r})=p_{\bot 0}({\bar r}) = 0. \nonumber
\end{eqnarray}
The two last equations correspond to the charged dust solution of the paper \cite{CooperstockDeLaCruz}, but in isotropic coordinates. For the charge distribution we have
\begin{equation}
l^2(\bar r)= \frac{16 {\bar r}^6}{a^2 \left( 1+{\bar r}^2 \right)^4}.
\end{equation}

\section{Results}
\label{results}

The integration of the differential equation (\ref{eqdiff}) depends on the coefficients $\bar{a}$ and $\bar{b}$ (equation (\ref{coeficientes})), related to the values of mass, radius and charge of the object before the beginning of the collapse. We have taken $m_0=5$M$_{\odot}$, $R=15$ km and different values for the ratio charge-mass $Q/m_0 \simeq 0.00$, $Q/m_0 = 0.20, 0.40, 0.64$. The result is shown in Figure \ref{f_carga}, from which we notice that the collapse is slower with the increasing of the charge.  This result is in contrast to that obtained by Medina et al. \cite{Medina88}, pointed as unexpected by the authors.

\begin{figure}[h]
\includegraphics[height=12cm,angle=-90]{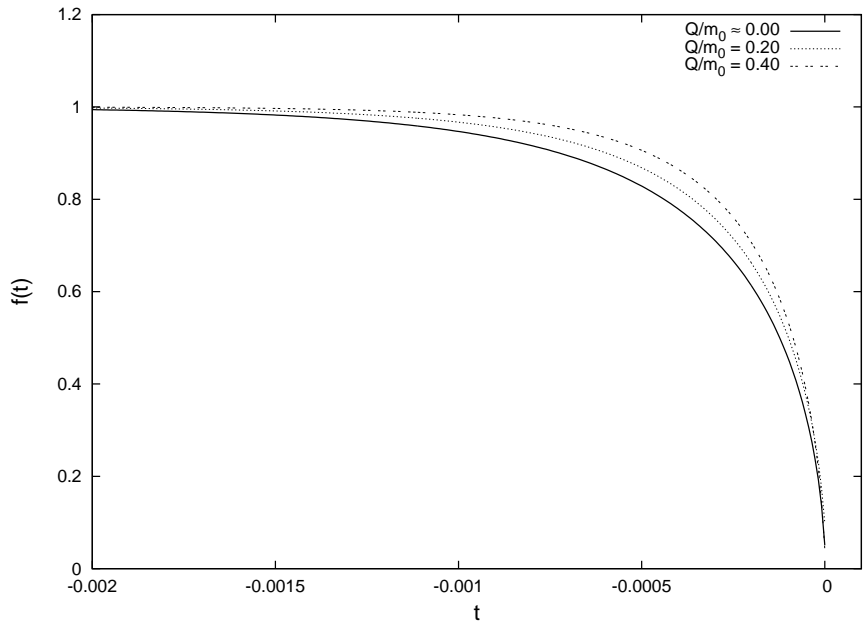}
\caption{\label{f_carga} Temporal behavior of the function $f(t)$ for the models 
with $Q/m_0\simeq0.0, 0.20, 0.40$. The time is in seconds.}
\end{figure}

In Section \ref{equations}, we declared that the formation of an event horizon 
occurs when the surface gravitational redshift goes to infinity, i.e., the term 
in the parentheses in the equation (\ref{eq:dvdtau1}) goes to zero. Using equations (\ref{eq:art0}), 
(\ref{eq:brt0}) and the metric functions (\ref{A0r}) and (\ref{B0r}) into the parentheses of that equation, we have
\begin{equation}
\left( 1 + {rB_{0}' \over B_0} + {r B_0 \dot f \over \ A_0} \right)_{\Sigma}=\left[ 1 -\frac{a^2R^2}
{1+\sqrt{1-a^2R^2}}+\frac{R \left(1+\sqrt{1-a^2R^2}\right)}{\beta} \dot{f_{bh}} \right]=0.
\label{eq:dvdtau2}
\end{equation}


So, the event horizon arises when
\begin{equation}
\dot{f}_{bh}=\frac{-\beta^2}{R \left( 1+\sqrt{1-a^2 R^2}\right)^2}.
\end{equation}

Note from Figure \ref{bhformationN} that, when the amount of charge is not very large, the event horizon is 
always formed, however, for larger values, this condition is no longer satisfied.

\begin{figure}[!h]
\begin{center}
\includegraphics[scale=0.35, angle=-90]{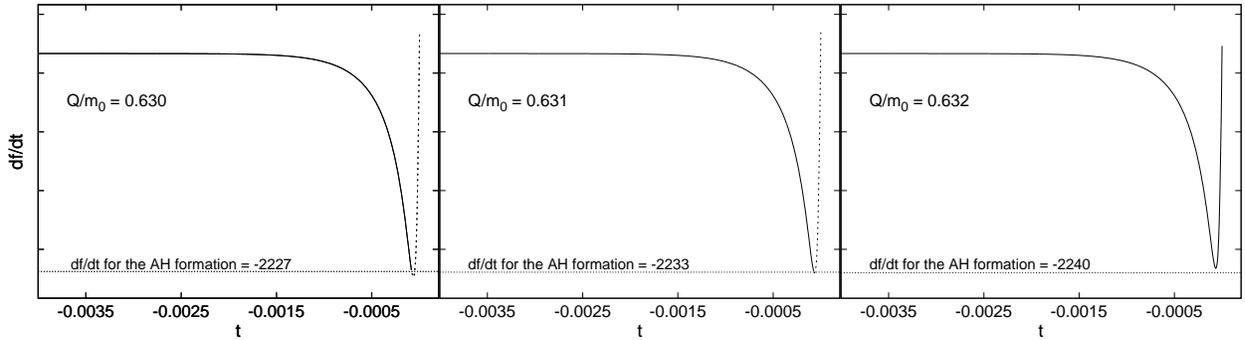}
\caption{\label{bhformationN} Gravitational redshift condition for a star 
with initial mass 5 M$\odot$ and initial radius $R=15$ km. For $Q/m_0 > 0.6310$, the function $\dot{f}$ no longer reaches the value for the event horizon formation.}
\end{center}
\end{figure}

In the last stages of the contraction of the overly charged stars ($Q/m_0 > 0.6310$), the decreasing 
of the function $f(t)$ decelerates and it reaches a minimum value (see the Figure \ref{inflexao}).
Beyond this point, the function increases sharply and a bounce of the system seems to occur. 
This idea is false, since the luminosity becomes negative for positive $\dot{f}$ (see equation (\ref{luminosidade}) below). This is physically inconsistent, and we have to stop the integration at this instant. 
The solution is just valid until this minimum point, where the luminosity, heat flux and the rate of decrease of the stellar radius ($\dot f$) go to zero. Nevertheless, the mass is reduced along the contraction by the radiative emission, but some portion still lasts at the endpoint. The same is truth for the stellar radius. 
This suggests that an equilibrium situation could be the final fate of the evolution.

\begin{figure}[!h]
\begin{center}
\includegraphics[scale=0.35, angle=-90]{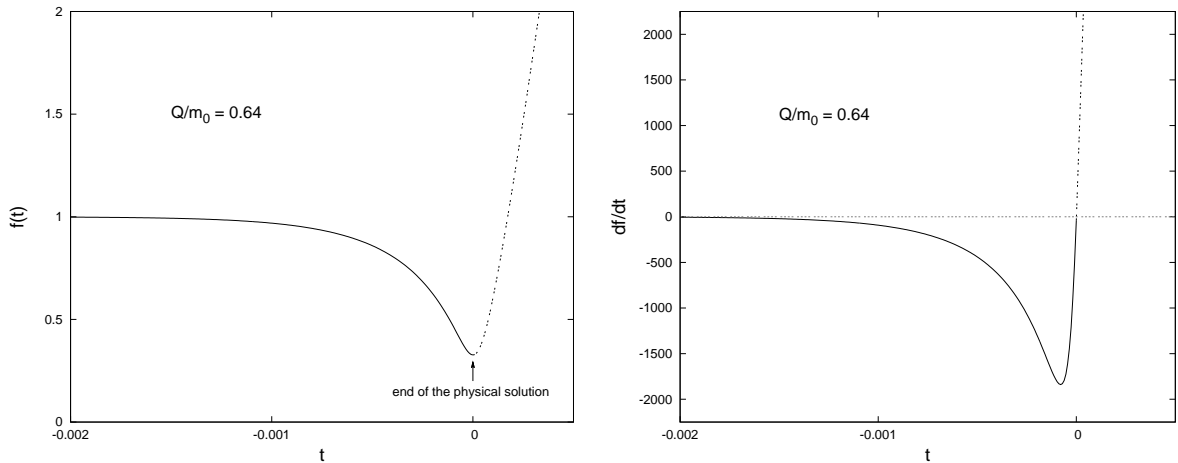}
\caption{\label{inflexao} Time evolution of $f(t)$ for a highly charged star.}
\end{center}
\end{figure}

The equations (\ref{eq:ms}) and (\ref{eq:lsp}) become 
\begin{equation}
m= \frac{R^3}{2 \beta^{2}} \left(1+\sqrt{1-a^2 R^2}\right) f \dot{f}^2 + 
\frac{a^2 R^3}{2} f + \frac{Q^2}{2Rf},
\label{massa}
\end{equation}

\begin{equation}
L= - \overline{a}R^2 \dot{f} \left[ 1 + \frac{R^2}{\beta^{2}} \left(2 \beta + 
a^2 R^2 \right) \dot{f} \right]^2.
\label{luminosidade}
\end{equation}

Note that, before the collapse, when $f(t) \rightarrow 1$ and 
$\dot{f} \rightarrow 0$, the equation (\ref{massa}) becomes
\begin{equation}
m_0= \frac{a^2 R^3}{2} + \frac{Q^2}{2R},
\label{massainicial}
\end{equation}
as we expected from \cite{CooperstockDeLaCruz}.

\begin{figure}[h]
\center
\subfigure{\includegraphics[scale=0.45, angle=-90]{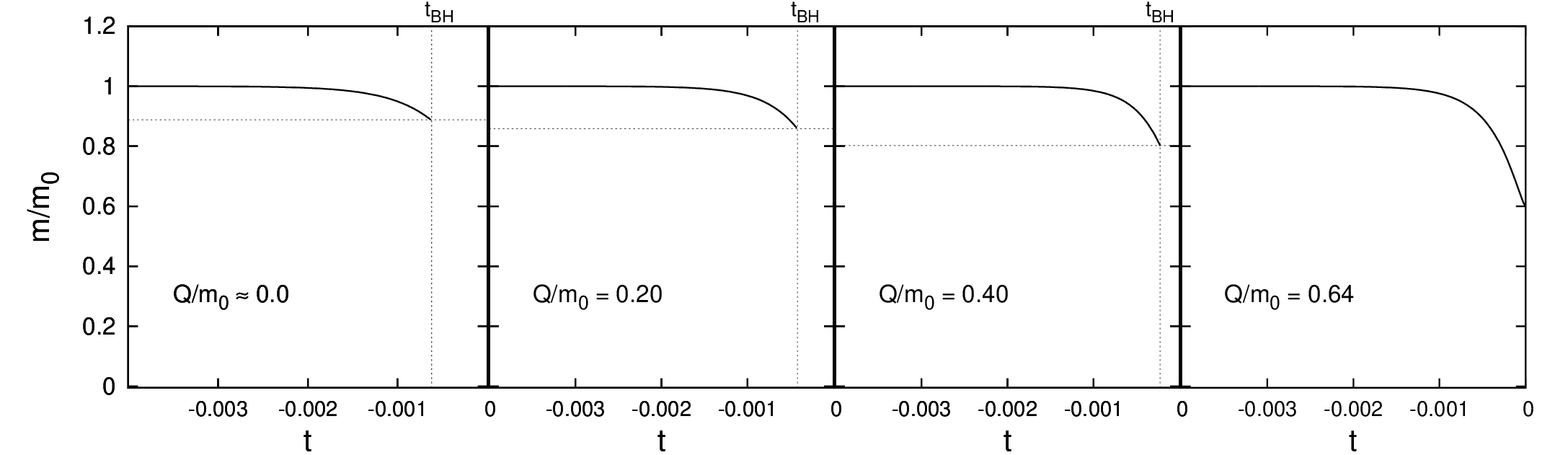}}
\subfigure{\includegraphics[scale=0.45, angle=-90]{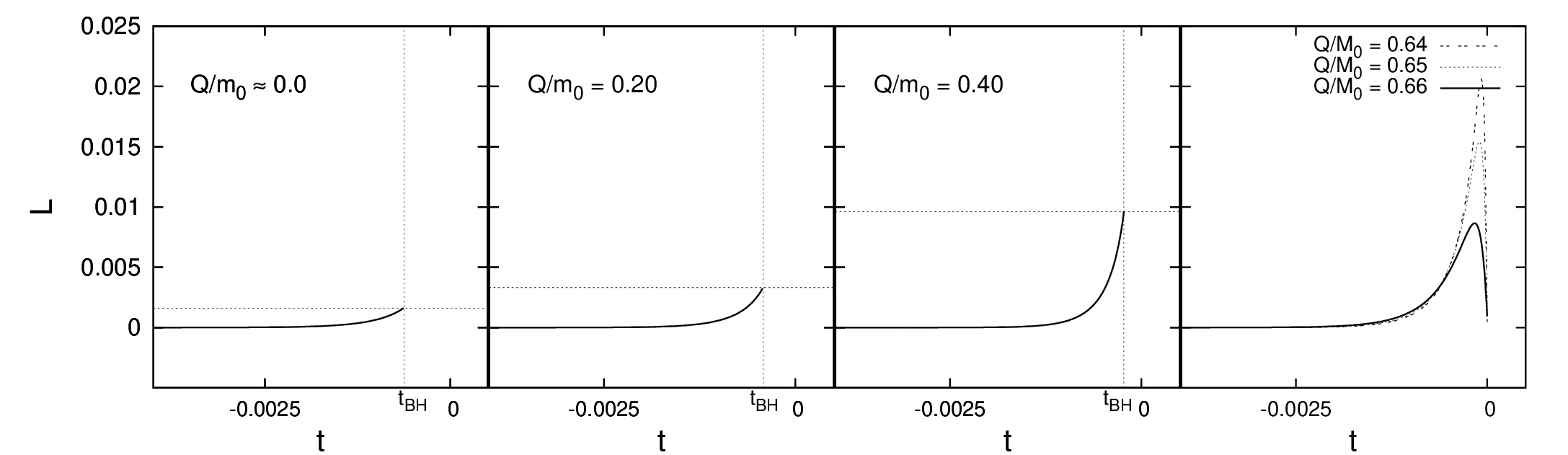}}
\caption{\label{massaluminosidade} Mass and luminosity for different 
charge/initial mass ratio models in geometric units.}
\end{figure}

The Figure \ref{massaluminosidade} shows the temporal evolution of the mass and luminosity. We can note that models with larger values of charge favor higher emission of energy. However, in the case where we have no black hole formation, this situation is reversed, as we can see from the right bottom panel.

The integrated physical quantities of the fluid are plotted in the figures \ref{densidade}, \ref{pressao}, \ref{pressaotang} and \ref{calor}. Due to the fact that the charge distribution is an increasing function of $r$, for the charged case, the radial pressure in the outermost radial coordinates are greater than the pressure in the innermost. 

\begin{figure}[!h]
\begin{center}
\includegraphics[scale=0.42, angle=-90]{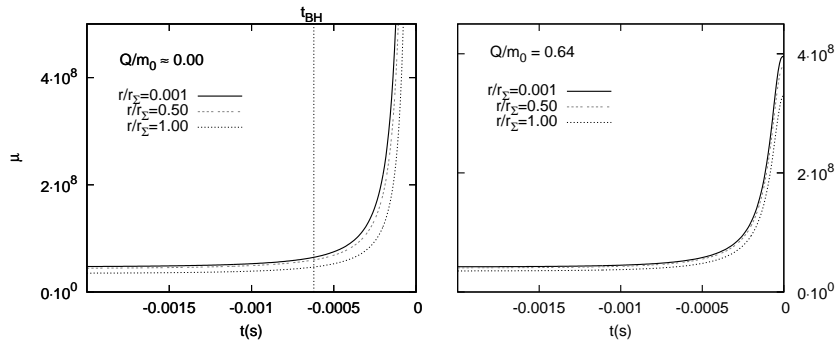}
\caption{\label{densidade} Temporal evolution of the density for an almost uncharged and a highly charged star in geometric units.}
\end{center}
\end{figure}

\begin{figure}[!h]
\begin{center}
\includegraphics[scale=0.4, angle=-90]{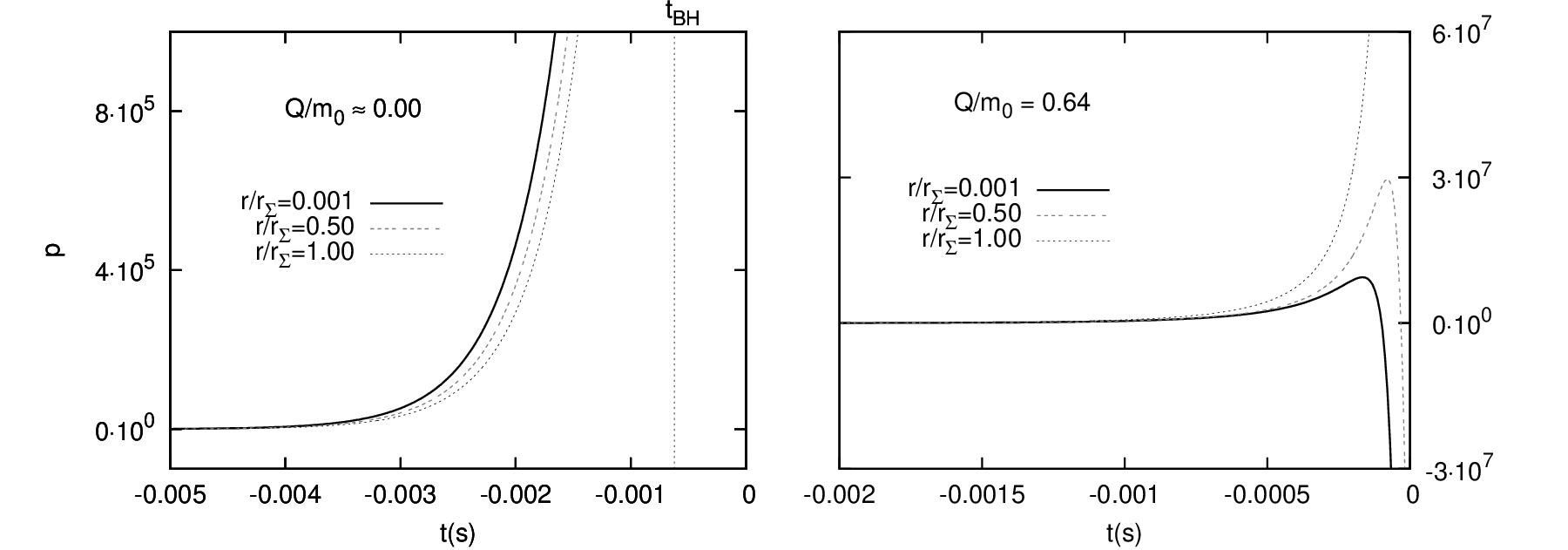}
\caption{\label{pressao} Temporal evolution of the pressure for an almost uncharged and a 
highly charged star in geometric units.}
\end{center}
\end{figure}

\begin{figure}[!h]
\begin{center}
\includegraphics[scale=0.4, angle=-90]{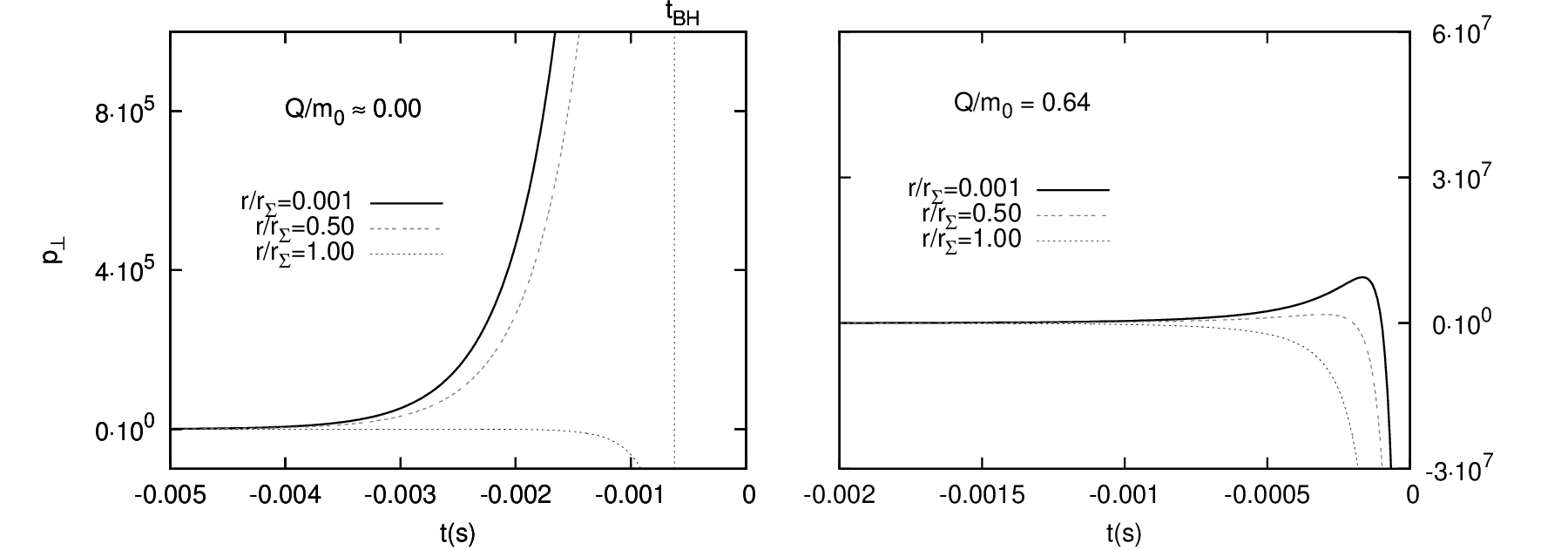}
\caption{\label{pressaotang} Temporal evolution of the tangential pressure for an almost uncharged and a 
highly charged star in geometric units.}
\end{center}
\end{figure}

\begin{figure}[!h]
\begin{center}
\includegraphics[scale=0.4, angle=-90]{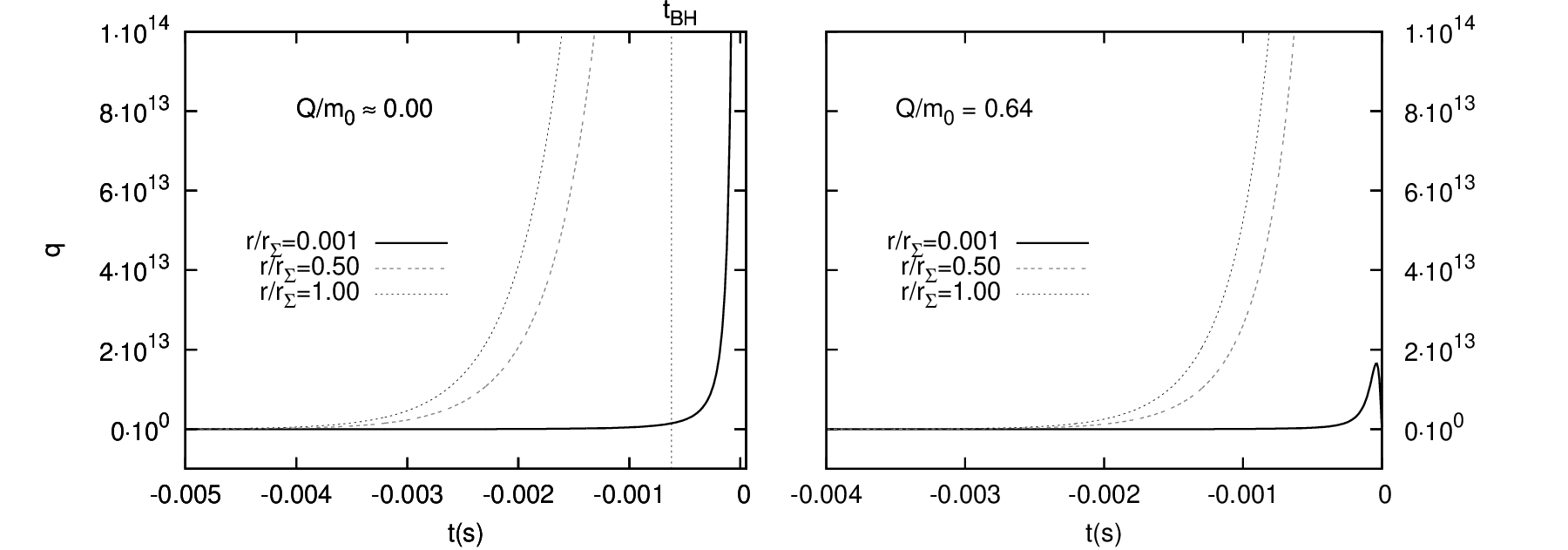}
\caption{\label{calor} Temporal evolution of the heat flux for an almost uncharged and a 
highly charged star in geometric units.}
\end{center}
\end{figure}

\section{Energy Conditions for The Charged Fluid}
\label{energyconditions}

In principle, any geometry might represent, via Einstein's equations, a valid solution of the theory of General Relativity, once nothing is inferred about the matter content of the source of the gravitational field, or, precisely, no restrictions are set on the components of the stress-energy tensor.
On the other hand, the known forms of matter are endowed with certain physical features, like positivity of the energy density or like its domain over the principal stresses, for example.
Desired physically meaningful solutions, these features are combined in a set of constraints, the energy conditions, guiding the acceptance or disposal of a particular solution of the field equations.

In order to trace the energy conditions along the moment of contraction of the body, we follow the same procedure 
used in Kolassis, Santos \& Tsoubelis \cite{Kolassis88}. Here we skip the mathematical step by step, warning that the similar approach is found in \cite{Chan00,Chan01}. 
These conditions for the charged fluid in question are fulfilled if the following inequalities are satisfied:

Mathematical condition

\begin{eqnarray}
(i)~~~~~~~~~~~~~~~\sqrt{\Delta} \geq 0 \nonumber
\end{eqnarray}

Common to all the energy conditions

\begin{eqnarray}
(ii)~~~~~~~~~~~\frac{1}{2} \left[ \mu + p + \frac{1}{8 \pi} \frac{{l^2(r)}}
{r^4 {B^{4}}} \left( \frac{1}{{B^2}} +2 \right) +\sqrt{\Delta} \right] \geq 0 \nonumber
\end{eqnarray}

Weak conditions:

\begin{eqnarray}
(iii)~~~~~~~~~~~~~~\frac{1}{2} \left[ \mu - p + \frac{1}{8 \pi} \frac{{l^2(r)}}
{r^4 {B^6}}  + \sqrt{\Delta} \right] \geq 0 \nonumber
\end{eqnarray}

Dominant conditions:

\begin{eqnarray}
(iv)~~~~~~~~\frac{1}{2} \left[ \mu - 3p + \frac{1}{8 \pi} \frac{{l^2(r)}}
{r^4 {B^4}} \left( \frac{1}{{B^2}} -2 \right)  + \sqrt{\Delta} \right] \geq 0 \nonumber
\end{eqnarray}

\begin{eqnarray}
(v)~~~~~~~~~~~~\frac{1}{2} \left[ \mu - p + \frac{1}{8 \pi} 
\frac{{l^2(r)}}{r^4 {B^6}} \right] \geq 0 \nonumber
\end{eqnarray}

Strong conditions:

\begin{eqnarray}
(vi)~~~~~~~~~~~~~\sqrt{\Delta} +2 p + \frac{1}{4 \pi} 
\frac{{l^2(r)}}{r^4 {B^4}}  \geq 0. \nonumber
\end{eqnarray}

Where

\begin{eqnarray}
\Delta = \left[ \left( \mu + p \right) - \frac{1}{8 \pi} 
\frac{{l^2(r)}}{r^4 {B^6}} \right]^2 - 4 {B^2} q^2. \nonumber
\end{eqnarray}

The Figure \ref{conditions064} shows that all these inequalities 
hold all along the dynamics of the charged fluid. 


\begin{figure}[h!]
\includegraphics[width=16cm,angle=-90]{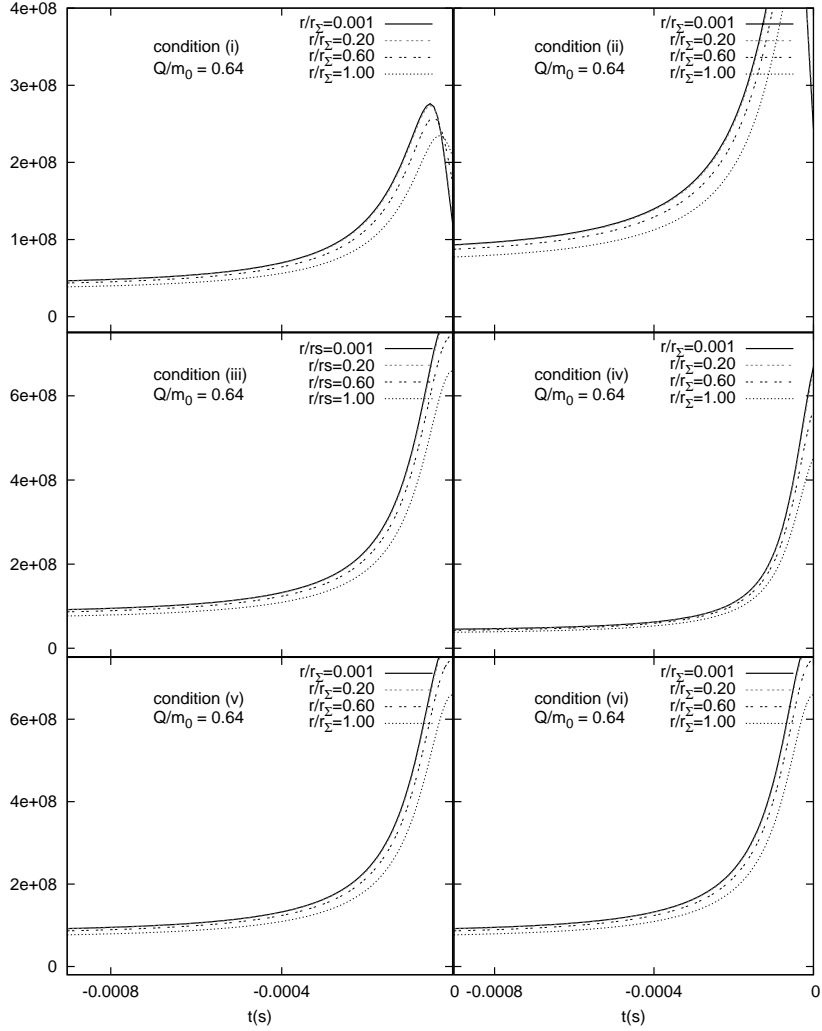}
\caption{\label{conditions064} Energy conditions for the highly charged model.}
\end{figure}

\newpage
\section{Discussion}
\label{discussion}

In this work we have studied the contraction of a spherically symmetric charged body. 
In order to make it possible the integration of the field equations, we have to adopt some
hypothesis about the interior spacetime. Firstly, its associated metric can be split into 
distinct functions of the radial and the temporal coordinates. Even though we can not guarantee the validity of this mathematical trick, it is a reasonable attempt to get round the underdetermination problem of the system of equations.
This drives us to the second assumption, the choice of the radial interior function. Among the uncountable possibilities of interior charged solutions, the Cooperstock and De La Cruz \cite{CooperstockDeLaCruz} solution was the selected one. Thus, the radial behavior of the physical variables of the system is an inheritance of such a particular choice.

In view of the previous discussion, we have concluded that appreciable physical effects to the collapse process are important for values of the net electric charge above $Q/m_0 \sim 0.10$ ($Q \sim 8.570 \times 10^{19}$ Coulomb). 
Furthermore, the Reissner-Nordstr\"om black hole is formed unless the total 
amount of charge is incredibly huge ($Q \sim 5.408 \times 10^{20}$ Coulomb for $Q/m_0 =0.631$).

The electric field delays the event horizon formation, in agreement with 
the non radiative models of \cite{Ghezzi05}, and can even prevent the complete contraction 
of the body. In this case, the solution seems to point to an equilibrium situation. 
It is even worthy to say that, for a model with ratio $Q/m_0 \gtrsim 0.70$, neither a minimal 
contraction takes place.

We also have seen that, for the models that admit black hole formation ($Q/m_0 \lesssim 0.631$), 
those richer in charge have greater transmition of energy to the exterior region 
before the appearance of the event horizon, making the remnant object less massive. 
The luminosity increases sharply and, all of a sudden, the star turns off. 
Otherwise, if it is charged enough in such a way to prevent the black hole formation, the time evolution of the luminosity has a maximum peak and a subsequent reduction to zero. In contrast, 
the event becomes less luminous with the increasing of the fraction $Q/m_0$.


Although the model presented here is, for sure, too much idealized, it certainly represents an interesting dynamical solution of Einstein's equation, and it gives us important clues on the possibility of the Reissner-Nordstr\"om black hole formation.

\noindent{\bf ACKNOWLEDGMENTS}

The author (RC) acknowledges the financial support from FAPERJ (no. E-26/171.754/2000, 
E-26/171.533/2002 and
E-26/170.951/2006) and from Conselho Nacional de Desenvolvimento Cient\'{\i}fico e 
Tecnol\'ogico - CNPq -
Brazil. The author (GP) acknowledges the financial support from CAPES, FAPERJ and CNPq.

\end{document}